\documentclass[11pt,a4paper]{article}

\usepackage[truedimen,margin=32mm]{geometry} 

\usepackage{mathrsfs}
\usepackage{amssymb}
\usepackage{amsmath}
\usepackage{ascmac}
\usepackage{amsthm}
\usepackage[pdftex]{graphicx}
\usepackage[pdftex]{color}
\usepackage{natbib}
\usepackage{setspace}
\usepackage{url}

\usepackage{color}

\usepackage{titlesec}
\titleformat*{\section}{\large\bfseries}
\titleformat*{\subsection}{\it}

\newtheorem{prp}{Proposition}

\def\th{{\theta}}

\def\hbeta{{\widehat{\beta}}}
\def\muh{{\widehat{\mu}}}
\def\thh{{\widehat{\th}}}
\def\sih{{\widehat{\sigma}}}

\title{{\bf On Selection Criteria for the Tuning Parameter in Robust Divergence }}

\date{}

\begin{document}

\maketitle
\doublespacing

\vspace{-1.5cm}
\begin{center}
{\large Shonosuke Sugasawa$^{1,3}$ and Shouto Yonekura$^{2,3}$}

\medskip

\medskip
\noindent
$^1$Center for Spatial Information Science, The University of Tokyo\\
$^2$Graduate School of Social Sciences, Chiba University\\
$^3$Nospare Inc.
\end{center}

\vspace{1cm}
\begin{center}
{\bf \large Abstract}
\end{center}
While robust divergence such as density power divergence and $\gamma$-divergence is helpful for robust statistical inference in the presence of outliers, the tuning parameter that controls the degree of robustness is chosen in a rule-of-thumb, which may lead to an inefficient inference.
We here propose a selection criterion based on an asymptotic approximation of the Hyvarinen score applied to an unnormalized model defined by robust divergence. 
The proposed selection criterion only requires first and second-order partial derivatives of an assumed density function with respect to observations, which can be easily computed regardless of the number of parameters. 
We demonstrate the usefulness of the proposed method via numerical studies using normal distributions and regularized linear regression.


\newpage
\section{Introduction}

Data with outliers naturally arise in diverse areas. In the analysis of data containing outliers, statistical models with robust divergence are known to be powerful and have been used regularly. In particular, the density power divergence \citep{basu1998robust} and $\gamma$-divergence \citep{fujisawa2008robust} have been routinely used in this context due to their robustness properties while there now exist others. Robust divergence, in general, holds a tuning parameter that controls robustness under model misspecification or contamination. \cite{basu1998robust} noted that there is a trade-off between estimation efficiency and strength of robustness; thereby, a suitable choice of the tuning parameter seems crucial in practice. However, a well-known selection strategy such as cross-validation is not straightforward under contamination, so that we need to rely on a trial-and-error way to find a reasonable value of the tuning parameter.

To select a turning parameter, we here propose a simple but novel selection criterion for the tuning parameter by using the asymptotic approximation of Hyvarinen score \citep{shao2019bayesian,dawid2015bayesian} with unnormalized models based on robust divergence.
Typical existing methods \citep{warwick2005choosing,basak2021optimal} choose a tuning parameter based on the asymptotic approximation of the mean square error but have the drawback of requiring some pilot estimators and an analytical expression of the asymptotic variance.
Besides, their works are essentially limited to the simple normal distribution and simple linear regression. 
Our proposed method has the following advantages over the existing studies.

\begin{enumerate}
    \item Our method does not require an explicit representation of the asymptotic variance. Therefore, our method can be applied to rather complex statistical models such as multivariate models, which seems difficult to be handled by the previous methods.
    \item In the existing studies, it is necessary to determine a certain value as a pilot estimate to optimize a tuning parameter. Thus, the estimates may strongly depend on the pilot estimate. On the other hand, our method does not require a pilot estimate and is stable and statistically efficient.
    \item Although our proposed method is based on a simple asymptotic expansion, it is more statistically meaningful and easier to interpret the results statistically than existing methods because it is based on the theory of parameter estimation for unnormalized statistical models.
\end{enumerate}

Through numerical studies under simple settings, we show that the existing methods can be sensitive to a pilot estimate and tends to select an unnecessarily larger value of a tuning parameter, leading to loss of efficiency compared with the proposed method. 
Moreover, we still apply the proposed selection method, an estimation procedure in which the asymptotic variance is difficult to compute. 
As an illustrative example of such a case, we consider robust linear regression with $\gamma$-divergence and $\ell_1$-regularization, where the existing approach is infeasible to apply.

As related works, there are two information criteria using the Hyvarinen score.
\cite{matsuda2019information} proposed AIC-type information criteria for unnormalized models by deriving an asymptotic unbiased estimator of the Hyvarinen score, but it does not allow unnormalized models whose normalizing constants do not exist. 
Hence, the criterion cannot be applied to the current situation. 
On the other hand, \cite{jewson2021general} proposed an information criterion via Laplace approximation of the marginal likelihood in which the potential function is constructed by the Hyvarinen score. 
Although \cite{jewson2021general} covers unnormalized models with possibly diverging normalizing constants, the estimator used in the criterion is entirely different from one defined as the maximizer of robust divergence; thereby, the criterion does not apply to the tuning parameter selection of robust divergence either.  
Moreover, \cite{yonekura2021adaptat} developed an robust sequential Monte Carlo sampler based on robust divergence in which $\gamma$ is adaptively selected. 
However, it does not provide selection of $\gamma$ in a frequentist framework.

The rest of the paper is organized as follows. 
Section \ref{sec:selection} introduces a new selection criterion based on the Hyvarinen score.
We then provide concrete expressions of the proposed criterion under density power divergence and $\gamma$-divergence in Section \ref{sec:div}.
We numerically illustrate the proposed method in two situations in Section \ref{sec:num}.
Concluding remarks are given in Section \ref{sec:conc}.

\section{Tuning parameter selection of robust divergence}
\label{sec:selection}

Suppose we observe $y_1,\ldots,y_n$ as realizations from a true distribution or data generating process $G$, and we want to fit a statistical model $\{f_{\theta}:\theta\in\Theta\}$ where $\Theta\subseteq\mathbb{R}^{d}$ for some $d\geq1$.
Further assume that the density of $G$ is expressed as $(1-\omega)f_{\theta^{\ast}}+\omega \delta$, where $\delta$ is a contaminated distribution that produces outliers in observations.
Our goal is to make statistical inference on $\theta^{\ast}$ by successfully eliminating information of outliers.  
To this end, robust divergence such as density power divergence \citep{basu1998robust} and $\gamma$-divergence \citep{fujisawa2008robust} is typically used for robust inference on $\theta^{\ast}$.
Let $y=(y_1,\ldots,y_n)$ be a vector of observations and $D_{\gamma}(y;\theta)$ be a (negative) robust divergence with a tuning parameter $\gamma$.
We assume that the robust divergence has a additive form, namely, $D_{\gamma}(y;\theta)=\sum_{i=1}^n D_{\gamma}(y_i;\theta)$, which are satisfied well-known robust divergences as discussed in Section \ref{sec:div}.

For selecting the tuning parameter $\gamma$, our main idea is to regard $L_{\gamma}(y_i;\theta)\equiv\exp\{D_{\gamma}(y_i;\theta)\}$ as an unnormalized statistical model whose normalizing constant may not exist. 
Recently, \cite{jewson2021general} pointed out that the role of such unnormalized models can be recognized in terms of relative probability. 
For such model, we employ the Hyvarinen score (H-score)  in terms of Bayesian model selection \citep{shao2019bayesian,dawid2015bayesian}, defined as
\begin{equation}\label{H-score}
H_n^{\ast}(\gamma)\equiv\frac1n\sum_{i=1}^n\left\{2\frac{\partial^2}{\partial y_i^2}\log L_{\gamma}^{(m)}(y)+\left(\frac{\partial}{\partial y_i}\log L_{\gamma}^{(m)}(y)\right)^2 \right\},
\end{equation}
where $L_{\gamma}^{(m)}(y)$ is the marginal likelihood given by 
\begin{equation}\label{ML}
L_{\gamma}^{(m)}(y)=\int \pi(\theta)\prod_{i=1}^n L_{\gamma}(y_i;\theta) d\theta.
\end{equation}
with some prior distribution $\pi(\theta)$.
We consider an asymptotic approximation of the H-score (\ref{H-score}) under large sample sizes.
Under some regularity conditions \citep[e.g.][]{geisser1990validity}, the Laplace approximation of (\ref{ML}) is 
\begin{equation}\label{laplace}
L_{\gamma}^{(m)}(y)\approx (2\pi)^{d/2}\pi(\thh_{\gamma})|H(\thh_{\gamma})|^{-1/2}\prod_{i=1}^n L_{\gamma}(y_i;\thh_{\gamma}),
\end{equation}
where $\thh_{\gamma}$ is the M-estimator given by 
$$
\thh_{\gamma}={\rm argmax}_{\theta} \sum_{i=1}^n \log L_{\gamma}(y_i;\theta),
$$
and $H(\thh_{\gamma})$ is the Hessian matrix at $\theta=\thh_{\gamma}$. 
Then, we have the following approximation, where the proof is deferred to Appendix.

\begin{prp}
\label{prop:approx}
Under some regularity conditions, it follows that 
\begin{align*}
\frac{\partial}{\partial y_i} \log L_{\gamma}^{(m)}(y)
=D_{\gamma}'(y_i;\thh_{\gamma})+o_p(1), \ \ \ \ 
\frac{\partial^2}{\partial y_i^2} \log L_{\gamma}^{(m)}(y)
=D_{\gamma}''(y_i;\thh_{\gamma})+o_p(1),
\end{align*}
where $D_{\gamma}'(y_i;\theta)=\partial D_{\gamma}(y_i;\theta)/\partial y_i$ and $D_{\gamma}''(y_i;\theta)=\partial^2 D_{\gamma}(y_i;\theta)/\partial y_i^2$.
\end{prp}

The above results give the following approximation of the original H-score: 
\begin{align}\label{criterion}
H_n(\gamma)
&=\frac1n\sum_{i=1}^n\left\{2D_{\gamma}''(y_i;\thh_{\gamma})+\left(D_{\gamma}'(y_i;\thh_{\gamma})\right)^2\right\}, 
\end{align}
which satisfies $H_n(\gamma)=H_n^{\ast}(\gamma)+o_p(1)$ under $n\to\infty$.
We then define the optimal $\gamma$ as 
$$
\gamma_{opt}={\rm argmin}_{\gamma}H_n(\gamma).
$$

Existing selection strategies for $\gamma$ mostly use the asymptotic variance of $\thh_{\gamma}$. 
For example, under the density power divergence, \cite{warwick2005choosing} and \cite{basak2021optimal} suggested using asymptotic approximation of the mean squared errors of $\thh_{\gamma}$.
However, computation of the asymptotic variance is not straightforward, especially when an additional penalty function is incorporated into the objective function or the dimension of $\theta$ is large.
On the other hand, the proposed criterion (\ref{criterion}) does not require the computation of asymptotic variance but only needs the derivatives of robust divergence concerning $y_i$.
Furthermore, it should be noted that the proposed criterion (\ref{criterion}) can be applied to a variety of robust divergence.

\section{Possible robust divergences to consider}
\label{sec:div}

We here provide detailed expressions for the proposed criterion (\ref{criterion}) under some robust divergences. 
For simplicity, we focus on two robust divergences which can be empirically estimated from the data. Still, the proposed method could be applied to other divergences such as Hellinger divergence \citep{devroye1985nonparametric} or $\alpha\beta$-divergence \citep{cichocki2011generalized}.  
In what follows, we shall use the notations, $f'(y_i;\th)=\partial f(y_i;\theta)/\partial y_i$ and $f''(y_i;\th)=\partial^2 f(y_i;\theta)/\partial y_i^2$.

\subsection{Density power divergence} 
The density power divergence \citep{basu1998robust} for a statistical model $f(y_i;\theta)$ is 
$$
D_{\gamma}(y_i;\theta)=\frac{1}{\gamma}f(y_i;\theta)^{\gamma}- \frac{1}{1+\gamma}\int f(t;\theta)^{1+\gamma}dt.
$$
It can be seen that $D_{\gamma}(y_i;\theta)+1-1/\gamma \to \log f(y_i;\theta)$ as $\gamma\to 0$, so the above function can be regarded as an extension of the standard log-likelihood. 
Then, a straightforward calculation leads to the expression of (\ref{criterion}), given by 
\begin{align*}
H_n(\gamma)
=\sum_{i=1}^n&\bigg[f'(y_i;\thh_{\gamma})^2f(y_i;\thh_{\gamma})^{\gamma-2}\left\{2(\gamma-1)+f(y_i;\thh_{\gamma})^{\gamma}\right\}+2f(y_i;\thh_{\gamma})^{\gamma-1}f''(y_i;\thh_{\gamma})\bigg].
\end{align*}

\subsection{$\gamma$-divergence} 
The original form of $\gamma$-divergence \citep{fujisawa2008robust} for a statistical model $f(y_i;\theta)$ is given by 
$$
\frac{1}{\gamma}\log\left\{\sum_{i=1}^n f(y_i;\theta)^{\gamma}\left(\int f(t;\theta)^{1+\gamma}dt\right)^{-\gamma/(1+\gamma)}\right\},
$$
which is not an additive form. 
However, the maximization of the above function with respect to $\theta$ is equivalent to the maximization of the transformed version of $\gamma$-divergence, $D_{\gamma}(y;\theta)=\sum_{i=1}^n D_{\gamma}(y_i;\theta)$, where 
$$
D_{\gamma}(y_i;\theta)=\frac{1}{\gamma}f(y_i;\theta)^{\gamma}\left\{\int f(t;\theta)^{1+\gamma}dt\right\}^{-\gamma/(1+\gamma)}.
$$
Then, we have 
\begin{align*}
H_n(\gamma)
=\sum_{i=1}^n&\bigg[f'(y_i;\thh_{\gamma})^2f(y_i;\thh_{\gamma})^{\gamma-2}\left\{\frac{2(\gamma-1)}{C_{\gamma}(\thh_{\gamma})}+\frac{f(y_i;\thh_{\gamma})^{\gamma}}{C_{\gamma}(\thh_{\gamma})^2}\right\}
+\frac{2f(y_i;\thh_{\gamma})^{\gamma-1}f''(y_i;\thh_{\gamma})}{C_{\gamma}(\thh_{\gamma})}\bigg],
\end{align*}
where $C_{\gamma}(\theta)=\left(\int f(t;\theta)^{1+\gamma}dt\right)^{\gamma/(1+\gamma)}$.

\section{Numerical examples}
\label{sec:num}

\subsection{Normal distribution with density power divergence}
We first consider a simple example of robust estimation of the normal population mean under unknown variance.
Let $y_1,\ldots,y_n$ be sampled observations and we fit $N(\mu,\sigma^2)$ to the data. 
The density power divergence of the model is given by 
\begin{equation*}
D_{\gamma}(y_i;\mu,\sigma^2)=\frac{1}{\gamma}\phi(y_i;\mu,\sigma^2)^{\gamma}-(2\pi\sigma^2)^{-\gamma/2}(1+\gamma)^{-3/2},    
\end{equation*}
where $\phi(y_i;\mu,\sigma^2)$ is the density function of $N(\mu,\sigma^2)$.
In this case, the criterion (\ref{criterion}) is expressed as 
$$
H_n(\gamma)
=\sum_{i=1}^n\left[\frac{2\left\{\gamma(y_i-\muh_{\gamma})^2-\sih_{\gamma}^2\right\}}{\sih_{\gamma}^4}\phi(y_i;\muh_{\gamma},\sih_{\gamma}^2)^{\gamma}
+
\frac{(y_i-\muh_{\gamma})^2}{\sih_{\gamma}^4}\phi(y_i;\muh_{\gamma},\sih_{\gamma}^2)^{2\gamma}\right],
$$
where $\muh_{\gamma}$ and $\sih_{\gamma}$ are the estimator based on the density power divergence.

We first demonstrate the proposed selection strategy through simulation studies. 
We simulated $y_1,\ldots,y_n$ from the normal distribution with true parameters, $\mu=2$, and $\sigma=1$, and then replace the first $n\omega$ observations by $y_i+7$.
We adopted four settings for $\omega\in \{0, 0.05, 0.1, 0.15\}$.
Using the simulated dataset, the optimal $\gamma$ is selected among $\{0, 0.01, \ldots,0.69, 0.70\}$ through the criterion $H_n(\gamma)$, and we obtain the adaptive estimator $\muh_{\gamma_{opt}}$.
For comparison, we also employed two selection methods, OWJ \citep{warwick2005choosing} and IWJ \citep{basak2021optimal}, in which the optimal value of $\gamma$ is selected via asymptotic approximation of mean squared errors of the estimator. 
We set $\gamma=0.5$ to compute a pilot estimator that must be specified in the two methods. 
Furthermore, we also computed $\muh_{\gamma}$ with $\gamma=0.1, 0.3$ and $0.5$.
Using an estimator of the asymptotic variance of $\muh_{\gamma}$ \citep[e.g.][]{basak2021optimal}, we also computed the Wald-type $95\%$ confidence interval of $\mu$.
Based on 5000 simulated datasets, we obtained the squared root of mean squared error (RMSE) of the point estimator as well as coverage probability (CP) and average length (AL) of the interval estimation. 
The results are reported in Table \ref{tab:sim-normal}.
It is observed that the use of small $\gamma$ (such as $\gamma=0.1$) may lead to unsatisfactory results when the contamination is heavy. 
It can also be seen that with the use of relatively large $\gamma$, the estimation results can be inefficient. 
On the other hand, the proposed method can adaptively select a suitable value of $\gamma$ as the averaged value of $\gamma_{opt}$ increases with the contamination ratio $\omega$, and it provides reasonable performance in all the scenarios.

\begin{table}[htbp!]
\caption{RMSE of the point estimation and CP and AL of interval estimation.   }
\label{tab:sim-normal}
\begin{center}
\begin{tabular}{ccccccccccc}
\hline
& & & & & & \multicolumn{3}{c}{fixed $\gamma$} \\
& $\omega$ &  & HS & OWJ & IWJ & $0.1$ & $0.3$ & $0.5$ \\
\hline
 & 0 &  & 10.3 & 10.6 & 10.3 & 10.2 & 10.5 & 11.0 \\
RMSE & 0.05 &  & 10.7 & 10.9 & 10.7 & 14.4 & 10.8 & 11.3 \\
 & 0.1 &  & 11.0 & 11.1 & 11.0 & 44.7 & 11.1 & 11.5 \\
 & 0.15 &  & 11.4 & 11.4 & 11.4 & 82.6 & 11.5 & 11.8 \\
 \hline
 & 0 &  & 94.8 & 93.8 & 94.2 & 94.6 & 94.5 & 94.4 \\
CP & 0.05 &  & 94.7 & 93.9 & 94.1 & 93.2 & 94.2 & 94.1 \\
 & 0.1 &  & 94.3 & 94.1 & 94.2 & 36.7 & 94.2 & 94.4 \\
 & 0.15 &  & 94.1 & 93.7 & 93.8 & 0.1 & 93.6 & 94.1 \\
 \hline
 & 0 &  & 40.6 & 40.1 & 39.8 & 40.4 & 40.7 & 42.6 \\
AL & 0.05 &  & 41.7 & 41.0 & 40.9 & 50.4 & 41.2 & 43.3 \\
 & 0.1 &  & 42.5 & 41.9 & 41.8 & 79.5 & 42.0 & 44.1 \\
 & 0.15 &  & 43.4 & 42.9 & 42.9 & 100.4 & 43.1 & 45.1 \\
\hline
\end{tabular}
\end{center}
\end{table}

\begin{table}[htbp!]
\caption{Average values of selected $\gamma$ in three methods.  }
\label{tab:sim-normal-gamma}
\begin{center}
\begin{tabular}{ccccccccccc}
\hline
$\omega$ &  & HS & OWJ & IWJ \\
\hline
0 &  & 0.088 & 0.212 & 0.158 \\
0.05 &  & 0.169 & 0.260 & 0.230 \\
0.1 &  & 0.217 & 0.284 & 0.267 \\
0.15 &  & 0.252 & 0.302 & 0.294 \\
\hline
\end{tabular}
\end{center}
\end{table}

\begin{figure}[!htb]
\centering
\includegraphics[width=7cm,clip]{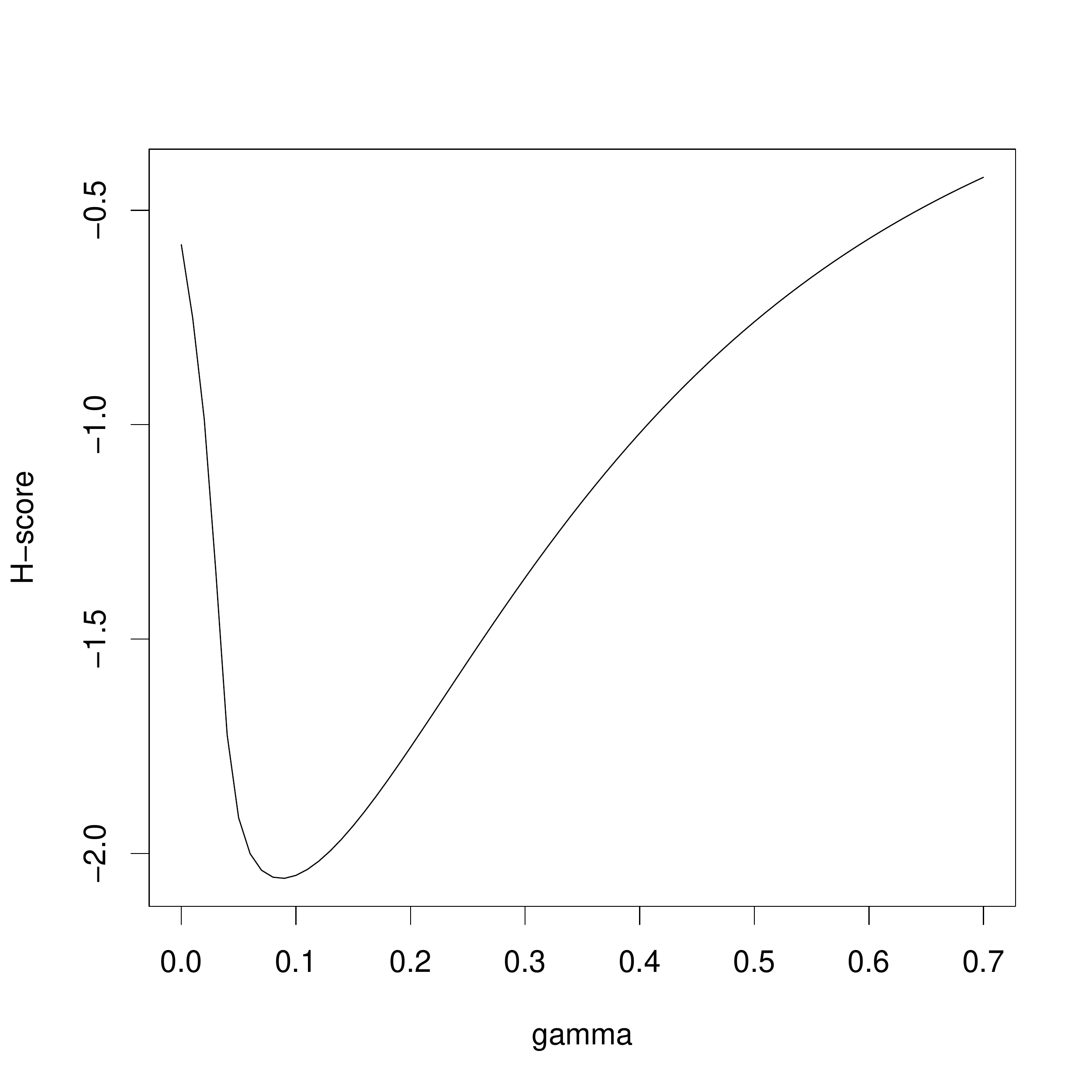}
\includegraphics[width=7cm,clip]{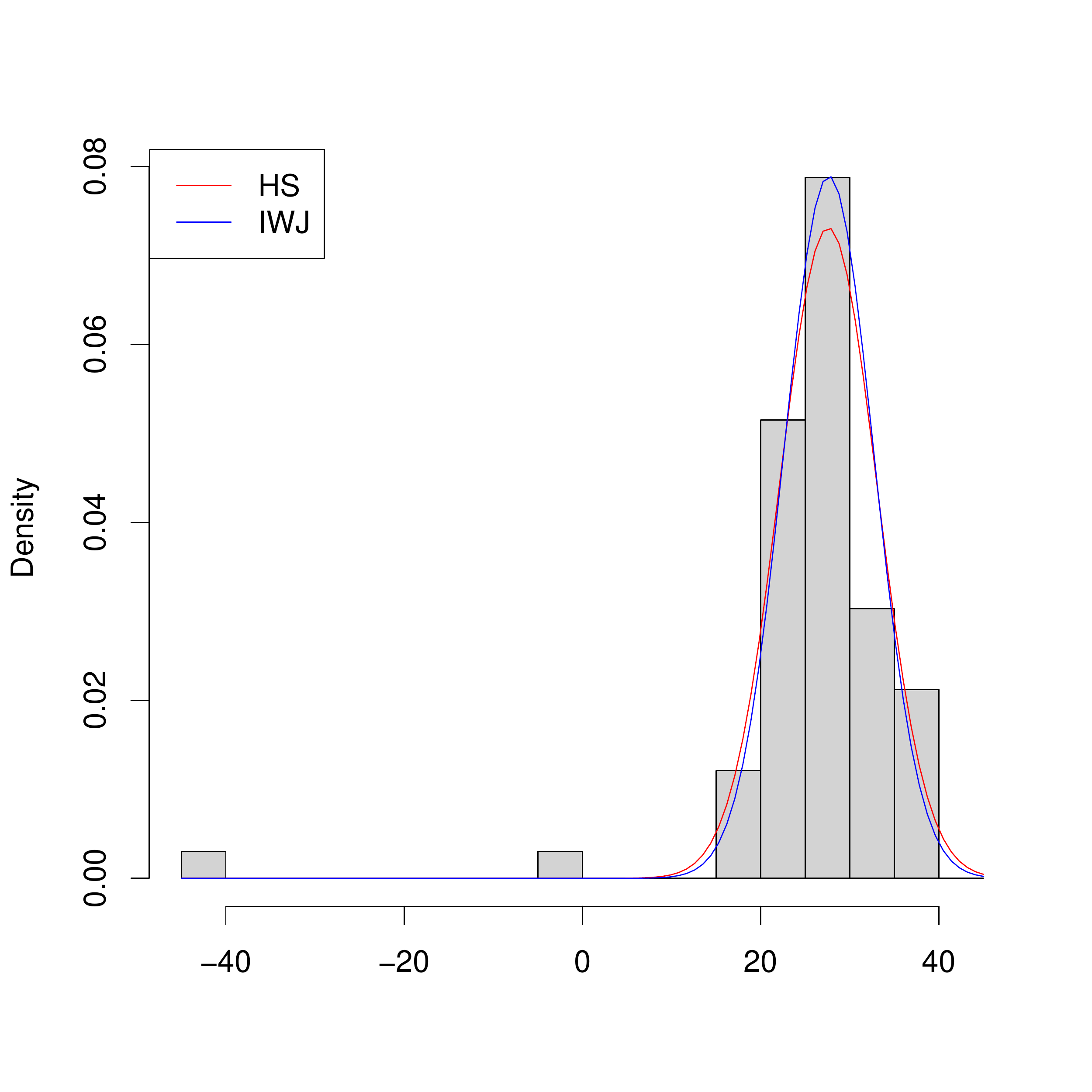}
\caption{H-scores for each $\gamma$ (left) and the estimate normal density functions with optimal gamma selected via the H-score and IJW methods (right).
\label{fig:Newcomb}
}
\end{figure}

We next apply the proposed method to Simon Newcomb's measurements of the speed of light data, motivated by applications in \citet{stigler1977robust,basu1998robust,basak2021optimal}.
We searched the optimal $\gamma$ among $\{0.01, 0.02,\ldots, 0.69, 0.70\}$ and the H-sores are shown in left panel in Figure \ref{fig:Newcomb}.
The obtained optimal value is $\gamma_{\rm opt}=0.09$, which is substantially smaller than $\hat{\gamma}=0.23$ selected by the existing methods as reported in \cite{basak2021optimal}.
Since the method proposed in \cite{basak2021optimal} requires a pilot estimate and the estimation results depend significantly on it, we believe that our estimation results are more reasonable. 
In fact, it is unlikely that we will have to use a value of  ${\gamma}=0.23$ for a data set that contains only two outliers. 
As shown in the right panel in Figure \ref{fig:Newcomb}, the estimated density functions are almost the same when $\gamma=0.09$ and when $\gamma=0.23$. However, it would be preferable to adopt the smaller value of $\gamma=0.09$ if the estimates are almost identical in terms of statistical efficiency.

\subsection{Regularized linear regression with $\gamma$-divergence}

Note that the proposed criterion can be used when some regularized terms are introduced in the objective function, while the existing method requiring an asymptotic variance of the estimator is not simply applicable.
We demonstrate the advantage of the proposed method through regularized linear regression with $\gamma$-divergence \citep{kawashima2017robust}.
Let $y_i$ and $x_i$ be a response variable and a $p$-dimensional vector of covariates, respectively, for $i=1,\ldots,n$. 
The model is $y_i\sim N(x_i^t\beta, \sigma^2)$.
Then, the transformed $\gamma$-divergence is  $D_{\gamma}(y_i;\theta)=\gamma^{-1}\phi(y_i;x_i^t\beta,\sigma^2)^{\gamma}/C_{\gamma}(\sigma^2)$ with $C_{\gamma}(\sigma^2)=\{(1+\gamma)^{-1/2}(2\pi\sigma^2)^{-\gamma/2}\}^{\gamma/(1+\gamma)}$, and the H-score is expressed as 
$$
H_n(\gamma)
=\sum_{i=1}^n\left[\frac{2\left\{\gamma(y_i-x_i^{\top}\hbeta_{\gamma})^2-\sih_{\gamma}^2\right\}}{\sih_{\gamma}^4 C_{\gamma}(\sih_{\gamma}^2)}\phi(y_i;x_i^{\top}\hbeta_{\gamma},\sih_{\gamma}^2)^{\gamma}
+
\frac{(y_i-x_i^{\top}\hbeta_{\gamma})^2}{\sih_{\gamma}^4C_{\gamma}(\sih_{\gamma}^2)^2}\phi(y_i;x_i^{\top}\hbeta_{\gamma},\sih_{\gamma}^2)^{2\gamma}\right].
$$
Here $\hbeta_{\gamma}$ and $\sih^2_{\gamma}$ are estimated as the minimizer of the following regularized $\gamma$-divergence:  
$$
-\frac{1}{\gamma}\log\left\{\sum_{i=1}^n \phi(y_i;x_i^{\top}\beta,\sigma^2)^{\gamma}\right\}-\frac{\gamma}{1+\gamma}\log \sigma^2 + \lambda \sum_{k=1}^p|\beta_k|,
$$
where $\lambda$ is an additional tuning parameter that can be optimized via 10-fold cross-validation. 
We use the R package \verb+gamreg+ \citep{kawashima2017robust} to estimate $\beta$ and $\sigma^2$ under given $\gamma$.

\begin{figure}[!htb]
\centering
\includegraphics[width=7cm,clip]{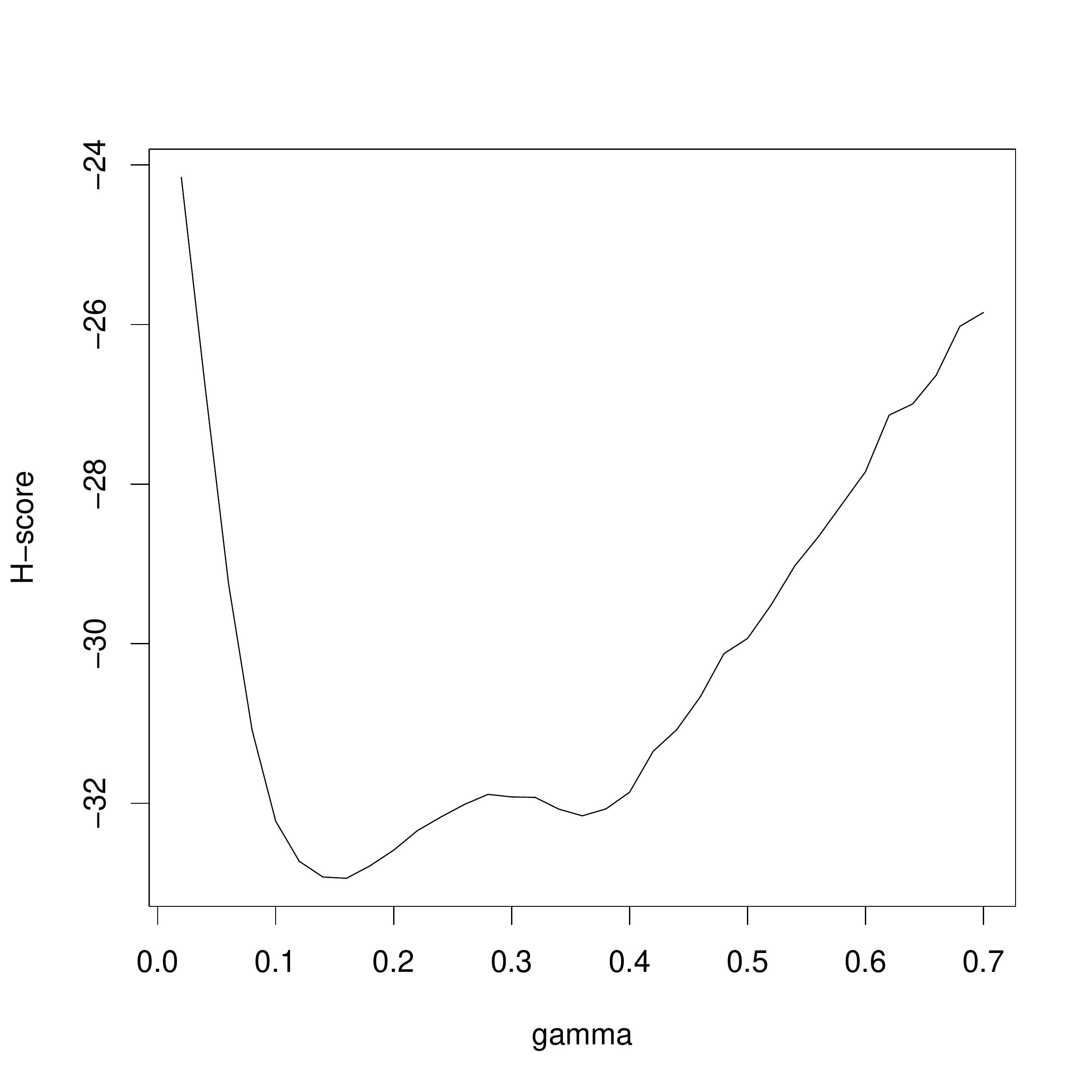}
\includegraphics[width=7cm,clip]{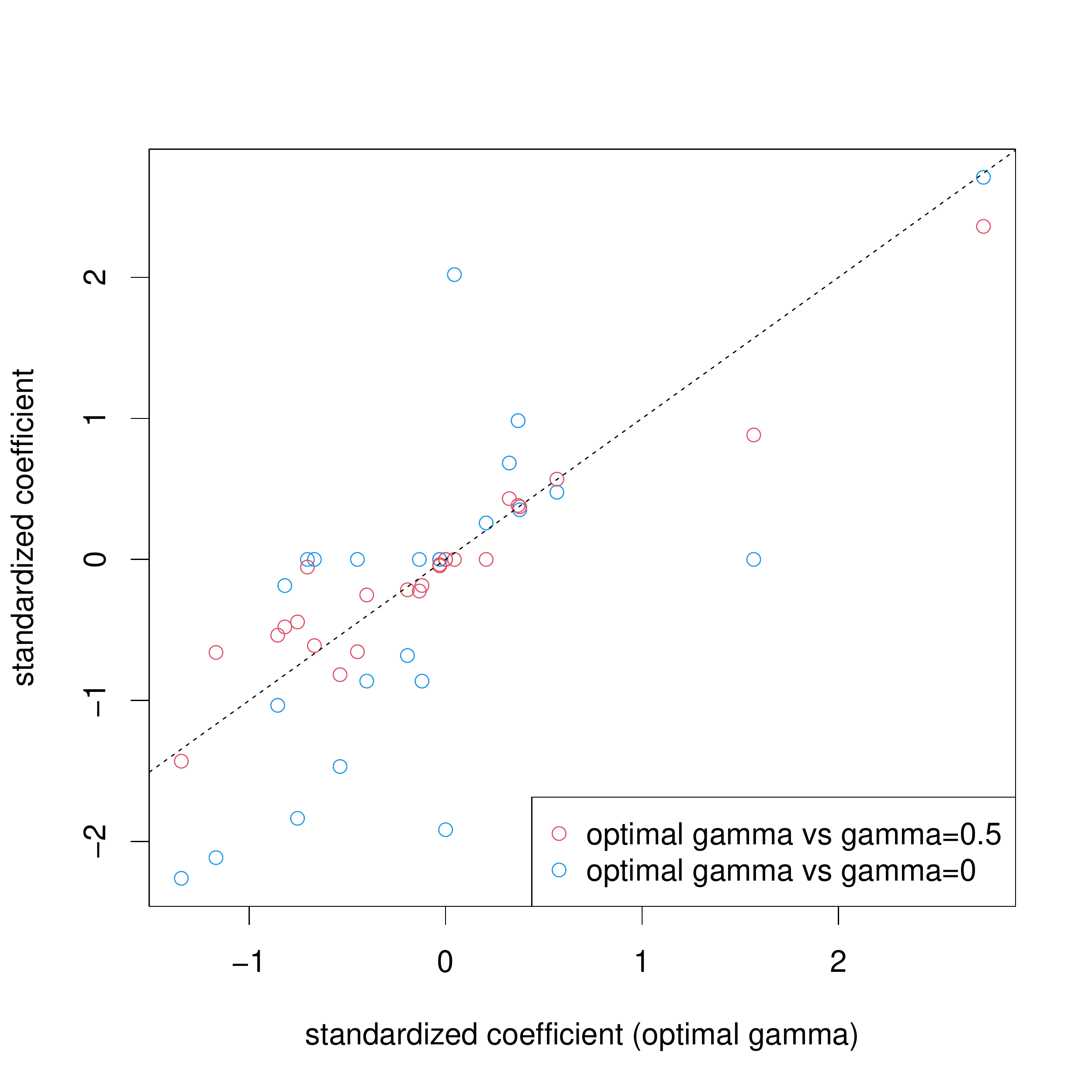}
\caption{H-scores for each $\gamma$ (left) and the estimated regression coefficients with three choices of $\gamma$ (right).
\label{fig:Boston}
}
\end{figure}

We apply the aforementioned method to the well-known Boston housing dataset \citep{harrison1978hedonic}.
In this analysis, we included the original 13 covariates and 12 quadratic terms of the covariates except for one binary covariate.
We searched the optimal $\gamma$ among $\{0.02, 0.04,\ldots, 0.68, 0.70\}$, and the estimated H-scores are shown in the left panel in Figure \ref{fig:Boston}, where the optimal value of $\gamma$ was $0.16$.
For comparison, we estimated the regression coefficients with $\gamma=0$ and $\gamma=0.5$. 
Note that $\gamma=0$ reduces to the (non-robust) standard regularized linear regression. 
The scatter plots of the estimated standardized coefficients under $\gamma=0.16$ against ones under the two choices of $\gamma$ are shown in the right panel of Figure \ref{fig:Boston}.
It is confirmed that the estimates with $\gamma=0.16$ and $\gamma=0.5$ are comparable while there are substantial differences between estimates with $\gamma=0.16$ and $\gamma=0$, indicating that a certain amount of robustness is required for the dataset.

\section{Concluding Remarks}
\label{sec:conc}

We proposed a new criterion for selecting the optimal tuning parameter in robust divergence, using the Hyvarinen score for unnormalized models with robust divergence. 
The proposed criterion does not require the asymptotic variance formula of the estimator that is needed in the existing selection methods.
Although we simply focused on the univariate and continuous situation, the proposed criterion can also be applied to multivariate or discrete distribution, where finite differences under discrete distributions should replace derivatives. 
Applications of the proposed score under such cases would also be helpful, and we left it to future work.

\vspace{1cm}
\begin{center}
{\bf Appendix}
\end{center}

\appendix
\section{Proof of Proposition \ref{prop:approx}}

We first assume standard regularity conditions in the M-estimation theory \citep[e.g.][]{van2000asymptotic} for the objective function $\sum_{i=1}^m \log L_{\gamma}(y_i;\theta)$. We also assume that $\log L_{\gamma}(y_i;\theta)$ is twice continuously differentiable with respect to $y_i$, $\log \pi(\theta)$ is continuously differentiable and the derivative of $\log \pi(\theta)$ is bounded.

We first note that $\thh_{\gamma}$ is a solution of the following estimating equation: 
$$
\sum_{i=1}^n S_{\gamma}(y_i;\theta)=0,  \ \ \ \ \ \  S_{\gamma}(y_i;\theta)\equiv \frac{\partial}{\partial\theta}\log L_{\gamma}(y_i;\theta).
$$
From the implicit function theorem, it follows that 
\begin{align*}
\frac{\partial\thh_{\gamma}}{\partial y_i}
&=\left\{\sum_{j=1}^n\frac{\partial}{\partial\theta}S_{\gamma}(y_j;\theta)\bigg|_{\theta=\thh_{\gamma}}\right\}^{-1}
\frac{\partial}{\partial y_i}\sum_{j=1}^n S_{\gamma}(y_j;\gamma)\bigg|_{\theta=\thh_{\gamma}}
=H(\thh_{\gamma})^{-1}S_{\gamma}'(y_i;\thh_{\gamma}),
\end{align*}
where we defined $S_{\gamma}'(y_i;\theta)=\partial S_{\gamma}(y_i;\theta)/\partial y_i$.
Note that $\partial\thh_{\gamma}/\partial y_i=O_p(n^{-1})$ under large $n$.
From (\ref{laplace}), the first order partial derivative of the marginal log-likelihood can be approximated as
\begin{equation}\label{log-approx}
\frac{\partial}{\partial y_i} \log L_{\gamma}^{(m)}(y)
\approx 
\frac{\partial}{\partial y_i} \sum_{j=1}^n \log L_{\gamma}(y_j;\thh_{\gamma}) + \frac{\partial}{\partial y_i}\log\pi(\thh_{\gamma})-\frac12\frac{\partial}{\partial y_i}\log |H(\thh_{\gamma})|.
\end{equation}
Under the regularity conditions for $\pi(\theta)$, it follows that 
$$
\frac{\partial}{\partial y_i}\log\pi(\thh_{\gamma})
=\frac{\partial\thh_{\gamma}}{\partial y_i}\times\frac{\partial}{\partial\theta}\log \pi(\theta)\bigg|_{\theta=\thh_{\gamma}} = o_p(1)
$$
under large $n$.
From the same argument, we can also show that $\partial\log |H(\thh_{\gamma})|/\partial y_i = o_p(1)$. 
Regarding the first term in (\ref{log-approx}), we have  
\begin{align}
\frac{\partial}{\partial y_i} \sum_{j=1}^n \log L_{\gamma}(y_j;\thh_{\gamma})
&=
\frac{\partial}{\partial y_i}\log L_{\gamma}(y_i;\theta)\bigg|_{\theta=\thh_{\gamma}}
+
\left(\frac{\partial\thh_{\gamma}}{\partial y_i}\right)^{\top}
\sum_{j=1}^n \frac{\partial}{\partial\theta}\log L_{\gamma}(y_j;\theta)\bigg|_{\theta=\thh_{\gamma}}
\notag \\
&=D'_{\gamma}(y_i;\thh_{\gamma})
+\left\{\sum_{j=1}^n S_{\gamma}(y_j;\thh_{\gamma})\right\}^{\top}
H(\thh_{\gamma})^{-1}
S_{\gamma}'(y_i;\thh_{\gamma})
\label{deriv-approx}\\
&=  D'_{\gamma}(y_i;\thh_{\gamma}) + O_p(n^{-1/2}), \notag
\end{align}
because $S_{\gamma}(y_j;\theta)$ is a score function and $\sum_{j=1}^n S_{\gamma}(y_j;\thh_{\gamma})=O_p(n^{1/2})$.

Using the expression of the first order derivative (\ref{deriv-approx}), it holds that 
\begin{align}
\frac{\partial^2}{\partial y_i^2} \sum_{j=1}^n \log L_{\gamma}(y_j;\thh_{\gamma}) 
&= 
\frac{\partial}{\partial y_i}D'_{\gamma}(y_i;\thh_{\gamma})
+
\left\{\frac{\partial}{\partial y_i}\sum_{j=1}^n S_{\gamma}(y_j;\thh_{\gamma})\right\}^{\top}
H(\thh_{\gamma})^{-1}
S_{\gamma}'(y_i;\thh_{\gamma}) \notag \\
& + 
\left\{\sum_{j=1}^n S_{\gamma}(y_j;\thh_{\gamma})\right\}^{\top}
H(\thh_{\gamma})^{-1}\left\{\frac{\partial}{\partial y_i}H(\thh_{\gamma})\right\}H(\thh_{\gamma})^{-1}
S_{\gamma}'(y_i;\thh_{\gamma}) \notag \\
& + 
\left\{\sum_{j=1}^n S_{\gamma}(y_j;\thh_{\gamma})\right\}^{\top}
H(\thh_{\gamma})^{-1}
\frac{\partial}{\partial y_i}S_{\gamma}'(y_i;\thh_{\gamma}). \label{deriv2-approx}
\end{align}
Note that 
$$
\frac{\partial}{\partial y_i}D'_{\gamma}(y_i;\thh_{\gamma})
= 
D''_{\gamma}(y_i;\thh_{\gamma}) + \left(\frac{\partial}{\partial \theta}D'(y_i;\theta)\bigg|_{\theta=\thh_{\gamma}}\right)^{\top}\frac{\partial\thh_{\gamma}}{\partial y_i}
=D''_{\gamma}(y_i;\thh_{\gamma}) + O_p(n^{-1}).
$$
By applying the same formula to $\partial S_{\gamma}'(y_i;\thh_{\gamma})/\partial y_i$, we can confirm that the third and forth terms in (\ref{deriv2-approx}) are $O_p(n^{-1/2})$.
Regarding the second term in (\ref{deriv2-approx}), we have 
\begin{align*}
\frac{\partial}{\partial y_i}\sum_{j=1}^n S_{\gamma}(y_j;\thh_{\gamma})  
&=S_{\gamma}'(y_i;\thh_{\gamma}) 
+
\left\{\sum_{j=1}^n \frac{\partial}{\partial\theta}S_{\gamma}(y_j;\theta)\bigg|_{\theta=\thh_{\gamma}}\right\}
H(\thh_{\gamma})^{-1}
S_{\gamma}'(y_i;\thh_{\gamma})\\
&=2S_{\gamma}'(y_i;\thh_{\gamma}),
\end{align*}
which shows that the second term in (\ref{deriv2-approx}) is $O_p(n^{-1})$, so that the proof is completed.

\vspace{1cm}
\bibliographystyle{chicago}
\bibliography{ref}

\end{document}